\documentclass[aps,reprint,prb,citeautoscript,superscriptaddress,amsmath,amssymb]{revtex4-2}
\usepackage{graphicx}
\usepackage{dcolumn}
\usepackage{bm}

\usepackage[version=4]{mhchem}
\usepackage[T1]{fontenc}
\usepackage[utf8]{inputenc}
\usepackage{multirow}
\usepackage{booktabs}
\usepackage{array}
\usepackage{tabularx}

\usepackage{xcolor}

\usepackage[normalem]{ulem}

\usepackage[percent]{overpic}

\usepackage{hyperref}
\usepackage{xr-hyper}  

\hypersetup{colorlinks=true,linkcolor=blue,filecolor=magenta,urlcolor=cyan}

\begin{document}

\title{Symmetry-selected extreme shear compliance in a stable carbon monolayer}

\date{\today}

\author{J. W. González}
\email[Corresponding author: ]{jhon.gonzalez@uantof.cl}
\affiliation{Departamento de Física, Universidad de Antofagasta, Av. Angamos 601, Casilla 170, Antofagasta, Chile}

\author{M. Pacheco}
\affiliation{Departamento de Física, Universidad Técnica Federico Santa María, Av. España 1680, Casilla 110V, Valparaíso, Chile}

\author{J. D. Correa}
\affiliation{Facultad de Ciencias Básicas, Universidad de Medellín, Medellín, Colombia}

\begin{abstract}
Extreme stiffness contrasts between loading modes usually require architected metamaterials with compliant hinges, whereas covalent crystals rarely exhibit such behavior intrinsically. Here, density functional theory calculations of the complete in-plane elastic tensors of four tetragonal carbon monolayers reveal a more than 50-fold spread in shear stiffness, despite less than a twofold variation in axial stiffness and the same symmetry-allowed tensor form. Twin T-graphene shows the strongest contrast: its shear constant, $C_{66}=3.03$~N\,m$^{-1}$, is 47 times smaller than in graphene and yields a shear Kelvin eigenvalue 29 times below the two normal-strain eigenvalues, while the Born stability criteria hold. This shear constant revises an earlier inference of elastic isotropy based on normal strains alone. Constrained relaxations and symmetry-resolved mode analysis identify the mechanism as a $B_{2g}$ counter-torsion of the carbon sublayers about the mid-plane bonds, which removes 96.3\% of the clamped-ion shear stiffness with bond-length changes below 0.15\%. An internal-coordinate model quantifies this cancellation, identifying internal relaxation, rather than weak bonding, as the origin of the soft channel. The resulting anisotropy predicts transverse sound velocities of 1.95~km\,s$^{-1}$ along [100] and 10.62~km\,s$^{-1}$ along [110], a testable acoustic signature. Phonon calculations, ab initio molecular dynamics at 300~K, and large-deformation simulations indicate that the soft channel is dynamically stable, preserves the local bonding, and persists beyond the harmonic regime. Our results show how bonding topology enables extreme mode-selective elasticity through internal relaxation in a mechanically stable covalent crystal.
\end{abstract}
\maketitle

\section{Introduction}

Materials that strongly resist some deformation modes while remaining exceptionally compliant in others show extremal elasticity. In two dimensions, one soft and two stiff elastic eigenvalues approach the unimode limit \cite{milton1995elasticity,milton2013complete}. Such responses occur mainly in architected metamaterials, where compliant hinges provide deformation pathways controlled by geometry \cite{kadic2012practicability,zhou2024two,czajkowski2024duality}. Crystalline solids soften through other mechanisms, including interlayer sliding in van der Waals materials such as graphite \cite{bosak2007elasticity}, directional anisotropy in puckered covalent monolayers such as black phosphorus \cite{pogna2022lattice,joshi2025lattice}, and switching between symmetry-related states in ferroelastic monolayers \cite{zhang2021flexible,wang2026metastable}. Whether a mechanically stable, densely bonded covalent crystal can isolate an exceptionally soft elastic mode through internal atomic relaxation alone remains an open question.

Internal relaxation may explain this behavior. Non-affine atomic displacements reduce elastic stiffness relative to the clamped-ion response, a phenomenon established in lattice theory \cite{born1954dynamical,martin1970elastic} and routinely incorporated in first-principles calculations \cite{wu2005systematic}. At the harmonic level, a strain component couples linearly only to internal modes of compatible symmetry so that the resulting softening can remain confined to a single elastic channel~\cite{gao2022anisotropic,rudenko2024anisotropic}. While symmetry dictates which relaxation channels are permitted, it does not determine their quantitative contributions. Extreme compliance requires a strain--mode coupling strong enough to nearly cancel the clamped-ion stiffness while keeping the remaining elastic eigenvalues positive. The separation of stiffness scales therefore depends on both the internal degrees of freedom and the bonding network that resists the other deformation modes.

\begin{figure*}[!]
\centering
\includegraphics[width=0.95\textwidth]{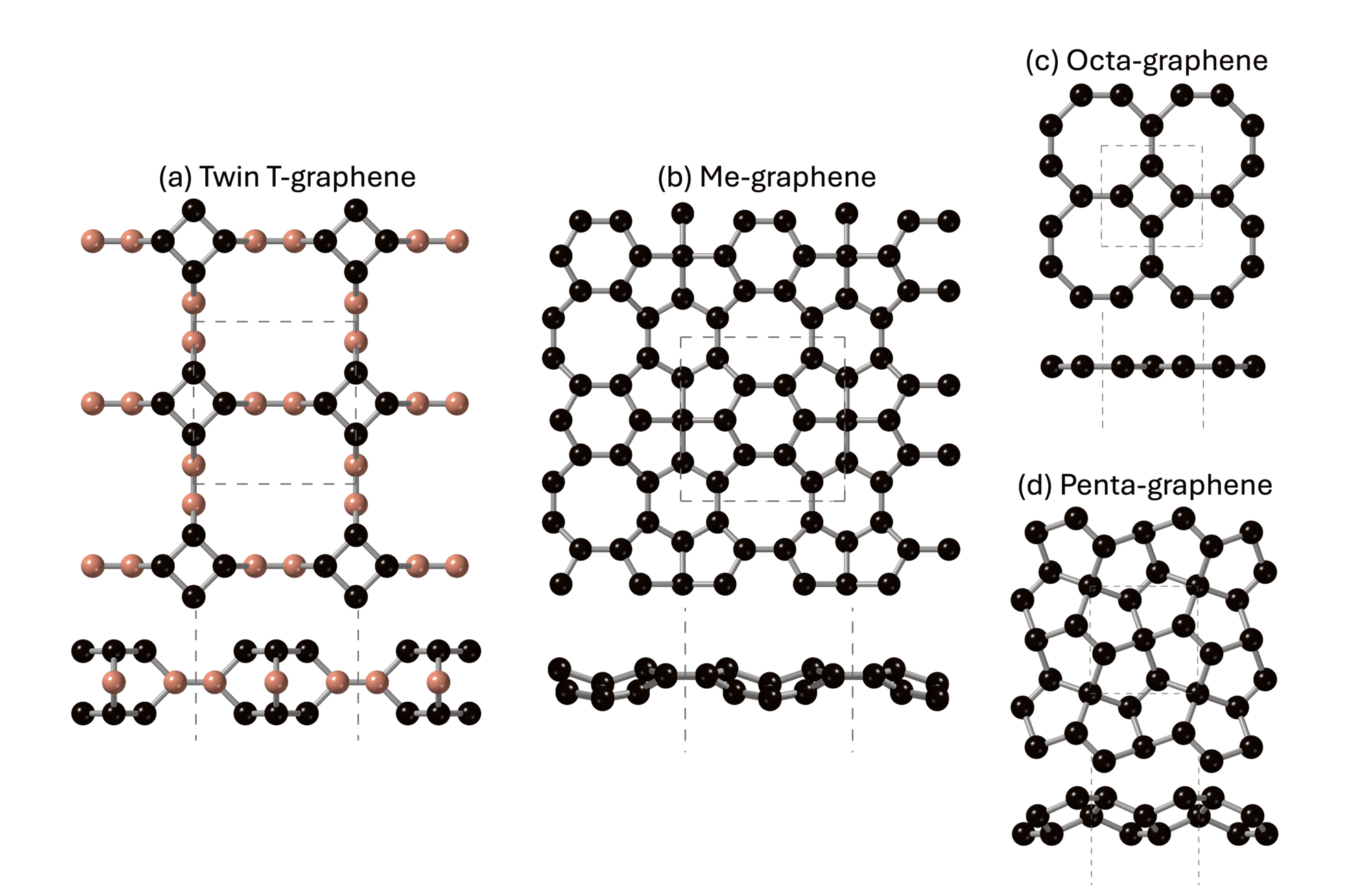}
\caption{
Optimized structures of \textbf{(a)} Twin T-graphene, \textbf{(b)}
Me-graphene, \textbf{(c)} octa-graphene, and \textbf{(d)}
penta-graphene, in top and side views; dashed lines mark the
primitive cells (structures not on a common scale). In Twin
T-graphene, black spheres are the threefold-coordinated
$\mathrm{C_A}$ atoms of the two outer tetragonal sublayers and pink
spheres are the mid-plane $\mathrm{C_B}$ atoms whose short
$\mathrm{C_B}$--$\mathrm{C_B}$ bonds act as the torsional axes
discussed in the text; the $\mathrm{C_B}$ atoms are held fixed
in the constrained-relaxation test.
}
\label{fig:scheme}
\end{figure*}

Such a compliant channel may evade conventional elastic characterization. Most studies of two-dimensional carbon allotropes beyond graphene focus on axial deformation, Young's moduli, and Poisson ratios \cite{sun2016new,ghosal2022beyond,zhao2013mechanical}. For a tetragonal monolayer, symmetry requires $C_{11}=C_{22}$, but full in-plane isotropy also demands $C_{66}=(C_{11}-C_{12})/2$. Calculations limited to principal-axis normal strains do not determine $C_{66}$ and therefore cannot confirm isotropy. In this work, we calculate the full in-plane elastic tensors of Twin T-graphene, Me-graphene, octa-graphene, and penta-graphene \cite{bhattacharya2020twin,zhuo2020me,sheng2012octagraphene,zhang2015penta}, four tetragonal carbon allotropes, using a consistent computational protocol, with graphene as an isotropic reference. Although these four allotropes share the same symmetry-allowed tensor form, they differ in bonding topology and sublayer geometry, allowing the comparison to isolate the influence of internal structure on the shear response under equivalent symmetry constraints.

The longitudinal stiffnesses of these allotropes differ by less than a factor of two, whereas their shear constants vary by more than a factor of fifty. Twin T-graphene shows the most pronounced contrast: $C_{66}=3.03$~N\,m$^{-1}$, 47 times smaller than in graphene, and a shear Kelvin eigenvalue 29 times lower than the two normal-strain eigenvalues, yet still satisfying the Born criteria. This shear constant revises the previous characterization of the monolayer as elastically isotropic, which rests solely on normal strains~\cite{bhattacharya2020twin}. Constrained relaxations and symmetry-resolved mode analysis demonstrate that shear couples to a $B_{2g}$ counter-torsion of the carbon sublayers about the short mid-plane bonds, removing 96.3\% of the clamped-ion shear stiffness with bond-length changes below 0.15\%. An internal-coordinate model quantitatively reproduces this cancellation. Transverse acoustic phonons independently validate the same soft channel and predict sound velocities of 1.95~km\,s$^{-1}$ along [100] and 10.62~km\,s$^{-1}$ along [110], a contrast measurable by Brillouin or inelastic scattering. The local bonding remains intact during ab initio molecular dynamics at 300~K over the simulated time, and large-deformation simulations using a machine-learning interatomic potential indicate that the compliant pathway persists beyond the harmonic regime.

These results show that internal relaxation, rather than weak bonding, drives the soft channel. They further indicate that bonding topology, rather than symmetry alone, determines whether such relaxation leads to extreme mode-selective elasticity. This mechanism suggests three criteria for identifying similar behavior in other crystals: the presence of an internal mode compatible with the applied strain, a coupling sufficiently strong to nearly cancel the corresponding clamped-ion stiffness, and a bonding network that maintains its stiffness under other deformation modes.

\section{Methodology}

\subsection{Computational details}
\label{sec:comp_details}

Density functional theory (DFT) calculations are performed with
\textsc{Quantum ESPRESSO} 7.4.1 \cite{giannozzi2009quantum,
giannozzi2020quantum} using the PBE functional
\cite{perdew1996generalized}, ultrasoft pseudopotentials from the efficiency
protocol of the SSSP library v2.0
\cite{prandini2018precision,Prandini2023SSSP}, kinetic-energy cutoffs of $60$~Ry for the wave-function 
and $500$~Ry for the density, $\Gamma$-centered grids of $9\times9\times1$ for relaxations and
$12\times12\times1$ for total energies, cold smearing
\cite{marzari1999thermal} of $0.002$~Ry, and DFT-D3(BJ) dispersion
\cite{grimme2010,grimme2011}. The structures are relaxed below
$10^{-3}$~Ry/bohr and $10^{-7}$~Ry, with more vacuum than $20$~\AA{}.
Harmonic force constants are obtained by finite displacements in a
$3\times3\times1$ supercell with \textsc{Phonopy}
\cite{togo2023implementation,togo2023first,phonopy-web}, enforcing the
acoustic sum rule.

The nonlinear large-deformation response is obtained with MACE-MH-1
\cite{batatia2025crosslearning,batatia2022mace} using the
\texttt{omat\_pbe} head \cite{barrosoluque2026omat24} through the Atomic
Simulation Environment \cite{larsen2017ase}, with a PBE-parametrized D3
correction \cite{grimme2010,grimme2011,takamoto2022}. All harmonic elastic
constants and all symmetry-resolved conclusions reported in the following are obtained
from DFT; MACE-MH-1 is used only to extend the analysis into the
large-deformation regime, after small-strain validation against the DFT
tensor.

\subsection{In-plane elastic tensor and directional response}
\label{sec:elastic_method}

The complete in-plane elastic tensor is extracted from total-energy
calculations under five homogeneous deformation families applied to 
optimized cells: uniaxial strain along $x$ and $y$, equibiaxial strain, simple
shear, and uniaxial strain along the lattice diagonal, with deformation
gradients
\begin{equation}
\begin{aligned}
\mathbf{F}_x &= \mathbf{I} + \lambda\,\mathbf{e}_x\otimes\mathbf{e}_x, &
\mathbf{F}_y &= \mathbf{I} + \lambda\,\mathbf{e}_y\otimes\mathbf{e}_y, \\
\mathbf{F}_{\mathrm{b}} &= (1+\lambda)\mathbf{I}, &
\mathbf{F}_{\mathrm{s}} &= \mathbf{I} + \gamma\,\mathbf{e}_x\otimes\mathbf{e}_y, \\
\mathbf{F}_{45} &= \mathbf{I} + \eta\,\mathbf{d}\otimes\mathbf{d}, &
\mathbf{d} &= \frac{\mathbf{e}_x+\mathbf{e}_y}{\sqrt{2}},
\end{aligned}
\label{eq:deformation_families}
\end{equation}
sketched in Fig.~S1 of the Supplemental Material. Each family uses
amplitudes $0$, $\pm0.02$, and $\pm0.04$ of its scalar parameter
($q=\lambda$, $\gamma$, or $\eta$), which cancels odd anharmonic
contributions and provides a residual-stress check. For every deformed cell
the lattice vectors are held fixed, and all internal coordinates relax fully,
so the uniaxial curvature returns $C_{11}$ directly rather than the Young's
modulus. Shear is applied as $x'=x+\gamma y$, so $\gamma$ is the engineering
shear strain, $\epsilon_{xy}=\gamma/2$, and the cell angle closes in an exactly
constant area; the diagonal family loads the normal and shear channels
simultaneously, making it an independent route to $C_{66}$.
The energy changes are normalized by the equilibrium area in the plane $A_0$, giving
two-dimensional constants in N\,m$^{-1}$
\cite{wei2009nonlinear,zhang2015penta,sun2016new}. In Twin T-graphene, the
shear scan costs almost two orders of magnitude less energy than the axial
one at equal amplitude and is therefore the numerically demanding family; it
sets the convergence thresholds quoted above, and the resulting fit quality is
documented in Sec.~S-I.

In Voigt notation, with
$\boldsymbol{\epsilon}=(\epsilon_{xx},\epsilon_{yy},\gamma_{xy})$ and 
engineering shear strain $\gamma_{xy}=2\epsilon_{xy}$, the harmonic energy
density is
$\Delta E/A_0=\tfrac{1}{2}\,\boldsymbol{\epsilon}^{\mathrm{T}}\mathbf{C}\,
\boldsymbol{\epsilon}$, with
\begin{equation}
\mathbf{C}
=
\begin{pmatrix}
C_{11} & C_{12} & 0 \\
C_{12} & C_{22} & 0 \\
0 & 0 & C_{66}
\end{pmatrix},
\label{eq:stiffness}
\end{equation}
where $C_{16}=C_{26}=0$ for the tetragonal classes of the
four monolayers. Each energy scan is fitted to a quadratic in its amplitude,
and the five curvatures yield $C_{11}$, $C_{22}$, $C_{12}$, and $C_{66}$
[Eqs.~(S1)--(S3)]. The simple-shear
scan provides the primary value $C_{66}^{\mathrm{shear}}$, and the diagonal
scan provides an independent estimate $C_{66}^{\mathrm{diag}}$. Mechanical stability
follows from the two-dimensional Born criteria,
\begin{equation}
C_{11}>0, \quad C_{22}>0, \quad C_{66}>0, \quad C_{11}C_{22}-C_{12}^{2}>0.
\label{eq:born}
\end{equation}

The extremal character of the tensor becomes explicit in the Kelvin (Mandel)
representation, in which the strain vector reads
$(\epsilon_{xx},\epsilon_{yy},\sqrt{2}\,\epsilon_{xy})$ and the stiffness
matrix carries the true tensor eigenvalues, 
\begin{equation}
\lambda_{\pm} = \frac{C_{11}+C_{22}}{2} \pm
\sqrt{\left(\frac{C_{11}-C_{22}}{2}\right)^{2} + C_{12}^{2}},
\quad \lambda_{\mathrm{sh}} = 2C_{66},
\label{eq:kelvin}
\end{equation}
which reduce under tetragonal symmetry to the compression, deviatoric, and
shear eigenvalues $\lambda_{\mathrm{comp}}=C_{11}+C_{12}$,
$\lambda_{\mathrm{dev}}=C_{11}-C_{12}$, and $\lambda_{\mathrm{sh}}=2C_{66}$.
Ordered as $\lambda_1\le\lambda_2\le\lambda_3$, the separation between
the soft and stiff eigenmodes is quantified by the near-unimode index
$\mathcal{R}_{u}=\lambda_2/\lambda_1$, which contrasts a unimode-like response
with the two-dimensional bimode limit of the Milton--Cherkaev classification
\cite{milton1995elasticity,zhou2024two}. The directional moduli $Y(\theta)$,
$G(\theta)$, and $\nu(\theta)$ follow from the rotation of the compliance tensor
$\mathbf{S}=\mathbf{C}^{-1}$ (Sec.~S-III), with shear
anisotropy $A_G=G_{\max}/G_{\min}$, and complete in-plane isotropy requires
\begin{equation}
C_{11}=C_{22}, \qquad C_{66} = \frac{C_{11}-C_{12}}{2},
\label{eq:isotropy}
\end{equation}
both of which are evaluated explicitly for every system. Finally, the internal
coordinate responsible for the soft shear channel is identified by projecting
the strain-induced non-affine relaxation onto the symmetry-allowed zone-center
optical modes and comparing it with the long-wavelength transverse acoustic
branch [Sec.~S-V, Eqs.~(S8)--(S9)].

\begin{table*}[t]
\centering
\caption{Optimized structural parameters of the four carbon allotropes, in the same
order as in Fig.~\ref{fig:scheme}. The structural prototypes were introduced
in Refs.~\protect\cite{bhattacharya2020twin,zhuo2020me,
sheng2012octagraphene,zhang2015penta}; all numerical values are from the
present calculations. The thickness is $h=z_{\max}-z_{\min}$ along the
monolayer normal, $N_z$ is the number of distinct atomic-height levels, and
$N_{\mathrm{b}}$ is the number of unique C--C bonds per primitive cell.
Detailed bond-length and bond-angle distributions are given in
Sec.~S-IV.}
\label{tab:geometry}
\begin{tabular}{lcccc}
\hline\hline
Property & Twin T-graphene & Me-graphene & Octa-graphene & Penta-graphene \\
\hline
Atoms per primitive cell, $N$ & 12 & 13 & 4 & 6 \\
Space group & $P4/mmm$ & $P\bar{4}m2$ & $P4/mmm$ & $P\bar{4}2_{1}m$ \\
Point group & $D_{4h}$ & $D_{2d}$     & $D_{4h}$ & $D_{2d}$ \\
$a=b$ (\AA) & 5.482 & 5.740 & 3.450 & 3.639 \\
In-plane area, $A_0$ (\AA$^2$) & 30.051 & 32.947 & 11.900 & 13.240 \\
Thickness, $h$ (\AA) & 2.107 & 1.002 & 0.000 & 1.220 \\
Atomic-height levels, $N_z$ & 3 & 5 & 1 & 3 \\
C--C bonds per cell, $N_{\mathrm{b}}$ & 18 & 20 & 6 & 10 \\
Bond-length range (\AA) & 1.332--1.478 & 1.405--1.564 & 1.372--1.469 & 1.339--1.552 \\
Bond-angle range ($^\circ$) & 90.0--134.5 & 95.9--142.6 & 90.0--135.0 & 98.9--133.7 \\
C atoms, threefold/fourfold & 12/0 & 12/1 & 4/0 & 4/2 \\
\hline\hline
\end{tabular}
\end{table*}

\section{Results}

\subsection{Geometric characterization}
\label{sec:geometry}

Figure~\ref{fig:scheme} shows the optimized monolayers and
Table~\ref{tab:geometry} shows their structural parameters. All four are
tetragonal, with a square metric, $a=b$ and a $90^{\circ}$ cell angle:
$D_{4h}$ for Twin T-graphene and octa-graphene, and $D_{2d}$ for Me-graphene
and penta-graphene. The first isotropy condition of
Eq.~\eqref{eq:isotropy} is therefore enforced by symmetry in all members of
the set, whereas the shear eigenvalue $2C_{66}$ remains an independent tensor
component in both point groups. Because the four monolayers share the same
symmetry-allowed form of the elastic tensor, symmetry alone does not
distinguish their shear responses. Any difference in $C_{66}$ instead reflects
the bonding topology, coordination, and out-of-plane morphology that control
the internal relaxation \cite{sun2016new,zhao2013mechanical}.

Twin T-graphene has the most pronounced out-of-plane architecture
\cite{bhattacharya2020twin}: three sublayers separated by $1.054$~\AA{} give
the largest thickness of the series, with tetragonal rings of $\mathrm{C_A}$
atoms in the outer sublayers, mid-plane $\mathrm{C_B}$ atoms bridging them,
and the shortest bonds of the structure, the $1.332$~\AA{}
$\mathrm{C_B}$--$\mathrm{C_B}$ links, lying entirely within the central plane.
This layered connectivity supplies the internal degree of freedom that becomes
decisive under shear.

\subsection{Geometric response under strain}
\label{sec:strain_response}

We next examine the structural response under the deformation families of
Sec.~\ref{sec:elastic_method}, evaluating the variations in bond-length and bond-angle
 in fixed connectivity. Under normal loading, all four allotropes
respond through the same local channels, with bond lengths changing by $2.65$
to $4.80\%$ and bond angles by $3.04^{\circ}$ to $10.72^{\circ}$, and the
corrugated structures couple their in-plane area to their thickness. Simple
shear separates the set into two classes. Penta-graphene, Me-graphene, and
octa-graphene accommodate $|\gamma|=0.04$ through those same channels, with
bond-length and bond-angle variations of $0.77$ to $1.44\%$ and $2.11^{\circ}$
to $2.95^{\circ}$, close to the affine bounds of $2.0\%$ and $2.29^{\circ}$.
Twin T-graphene responds qualitatively differently: at the same shear strain, its bond lengths vary by at most $0.14\%$, its bond angles by at most
$0.59^{\circ}$, and its thickness by $\sim0.001$~\AA. Thus, bond stretching, angle
bending, and out-of-plane relaxation are all suppressed, and this does
not follow from thickness or corrugation alone, since Twin T-graphene is the
thickest member of the set yet shows the smallest local distortions while
penta-graphene and Me-graphene combine several height levels without a
comparable response (complete statistics in Sec.~S-IV).
If the covalent bonds barely distort, what accommodates the shear?

\begin{table*}[t]
\centering
\caption{
In-plane elastic constants, in N\,m$^{-1}$, and derived directional
properties, all obtained from the present calculations.
$C_{66}^{\mathrm{shear}}$ follows directly from the simple-shear energy scan,
whereas $C_{66}^{\mathrm{diag}}$ is inferred independently from uniaxial
deformation along the $45^{\circ}$ lattice diagonal. Here $Y(\theta)$ is the
directional Young's modulus, $\nu_{xy}=C_{12}/C_{11}$ the axial Poisson's
ratio, and $A_G=G_{\max}/G_{\min}$ the shear-anisotropy factor. For
Me-graphene, tetragonal symmetry imposes $C_{11}=C_{22}$, and the two
values obtained independently from the $x$ and $y$ uniaxial scans agree to
within $10^{-2}$~N\,m$^{-1}$, which provides a direct measure of the numerical
accuracy of the curvature fits.
Graphene serves as an
isotropic control, for which hexagonal symmetry imposes
$C_{66}=(C_{11}-C_{12})/2$ and $A_G=1$.
}
\label{tab:elastic}
\begin{tabular}{lccccccccc}
\hline\hline
System
& $C_{11}$
& $C_{12}$
& $C_{66}^{\mathrm{shear}}$
& $C_{66}^{\mathrm{diag}}$
& $\dfrac{C_{11}-C_{12}}{2}$
& $Y(0^\circ)$
& $Y(45^\circ)$
& $\nu_{xy}$
& $A_G$ \\
\hline

Twin T-graphene
& 178.2 & $-1.40$ & 3.03  & 3.20  & 89.8
& 178.1 & 11.7  & $-0.008$ & 29.7 \\

Me-graphene
& 207.3 & 3.36 & 103.4 & 102.2 & 102.0
& 207.3 & 208.7 & 0.016 & 1.0 \\

Octa-graphene
& 299.9 & 55.55 & 48.6 & 47.7 & 122.2
& 289.6 & 152.6 & 0.185 & 2.5 \\

Penta-graphene
& 265.4 & $-17.16$ & 152.1 & 151.8 & 141.3
& 264.3 & 273.4 & $-0.065$ & 1.1 \\

Graphene
& 357.1 & 66.78 & 142.8 & 153.0 & 145.1
& 344.5 & 344.5 & 0.187 & 1.0 \\

\hline\hline
\end{tabular}
\end{table*}

\begin{figure*}[t]
\centering
\includegraphics[width=0.32\textwidth]{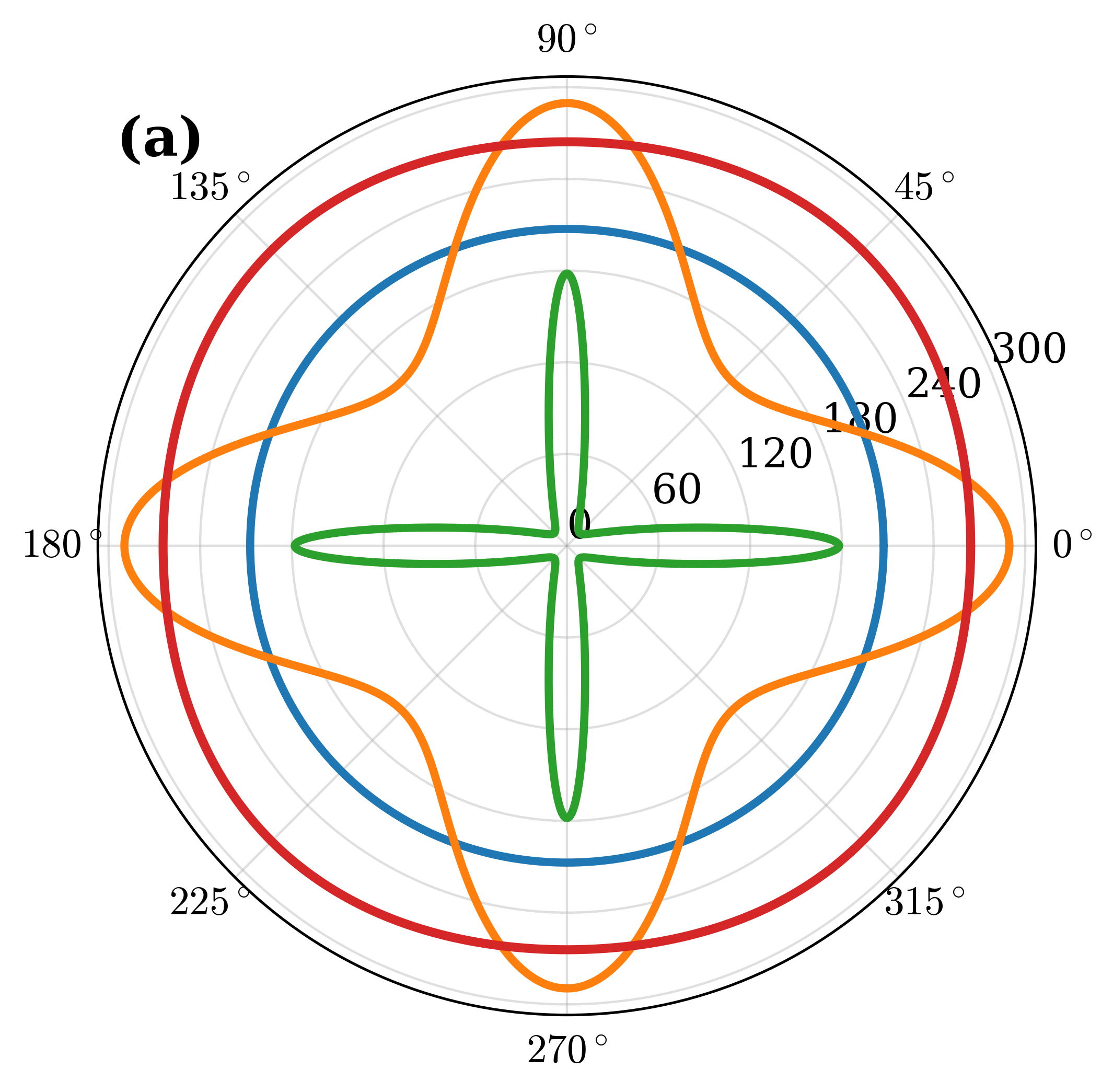}\hfill
\includegraphics[width=0.32\textwidth]{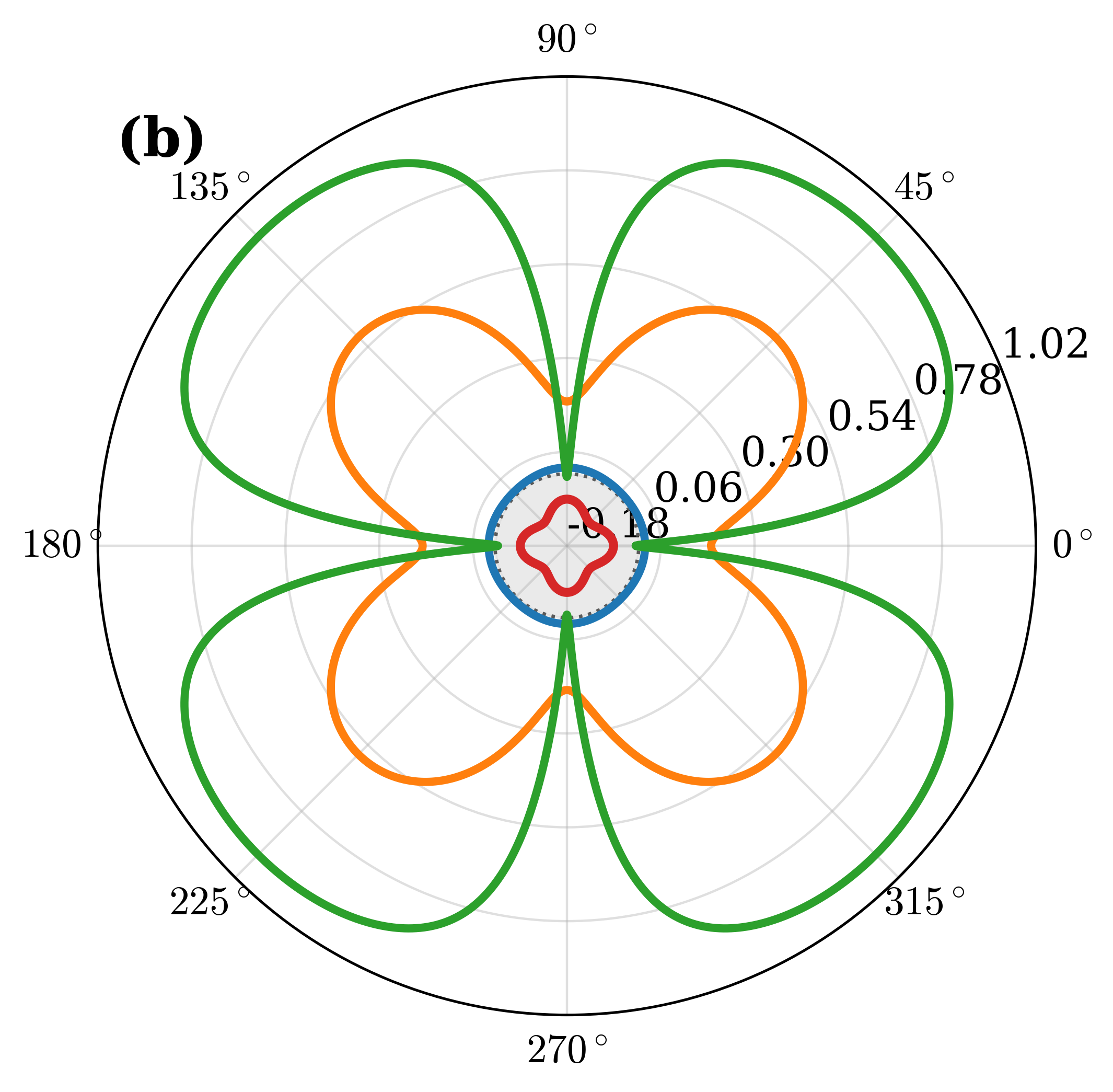}\hfill
\includegraphics[width=0.32\textwidth]{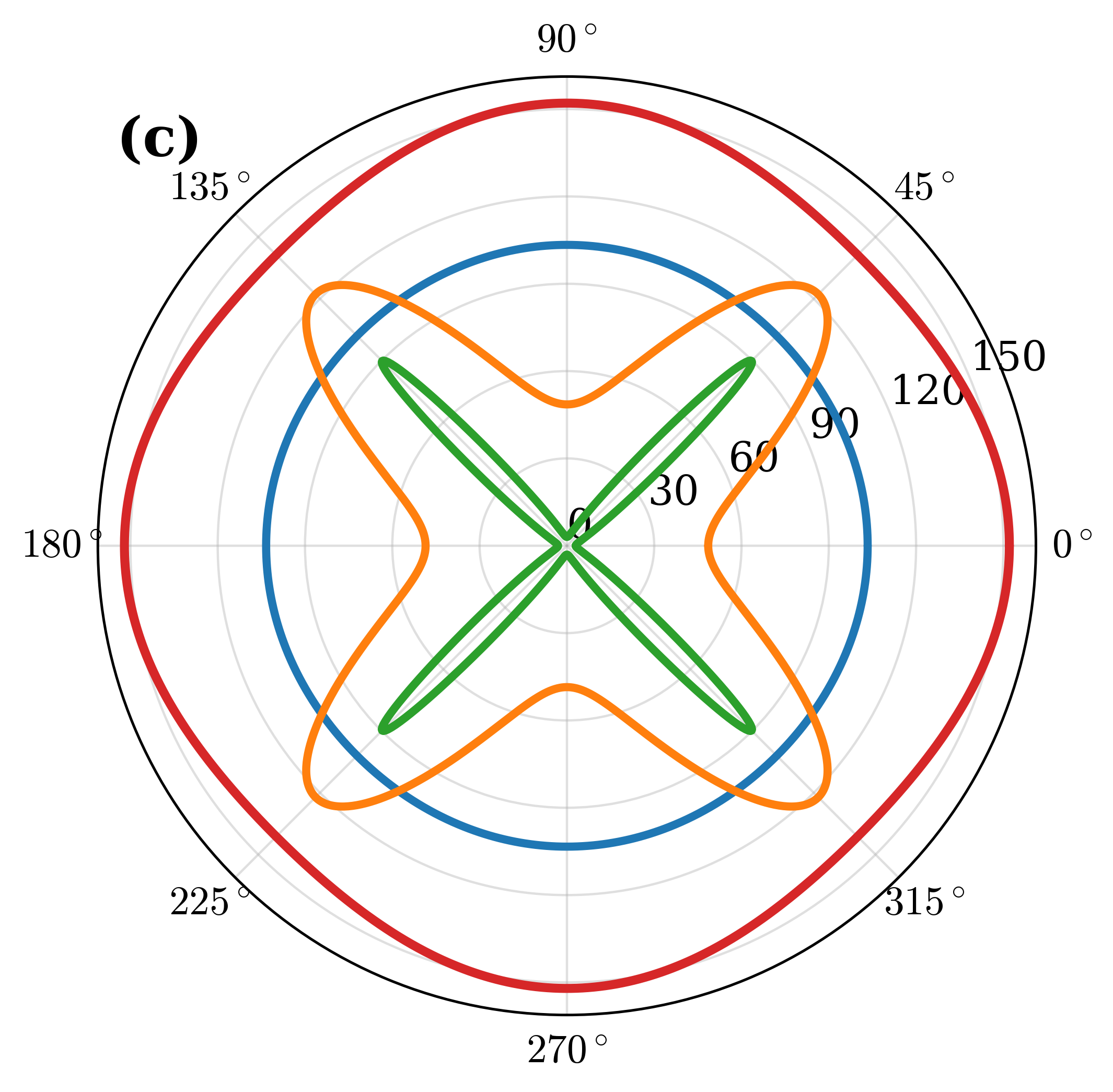}
\caption{
Directional in-plane elastic properties of the four carbon monolayers:
\textbf{(a)} Young's modulus $Y(\theta)$, \textbf{(b)} Poisson's
ratio $\nu(\theta)$, and \textbf{(c)} shear modulus $G(\theta)$, with
moduli in N\,m$^{-1}$ (penta-graphene red, Me-graphene blue,
octa-graphene orange, Twin T-graphene green). The dotted circle in
\textbf{(b)} marks $\nu=0$; a circular curve indicates an isotropic
response.
}
\label{fig:polar}
\end{figure*}

\subsection{Elastic constants, Kelvin spectrum, and in-plane anisotropy}
\label{sec:elastic}

Table~\ref{tab:elastic} summarizes the elastic constants and directional
moduli. Our axial response reproduces the published characterization of Twin
T-graphene, $C_{11}=178.2$~N\,m$^{-1}$ against $177.25$~N\,m$^{-1}$
\cite{bhattacharya2020twin}, and the corresponding benchmarks for the
remaining systems (Sec.~S-II); what that characterization lacks
is the shear component. Two independent deformation routes supply it and give
$C_{66}=3.03$ and $3.20$~N\,m$^{-1}$, agreeing to within the accuracy the same
two routes display on graphene, where symmetry fixes the exact answer
(Sec.~S-I). The diagonal route obtains $C_{66}$ by subtracting the
normal-strain contribution $(C_{11}+2C_{12}+C_{22})/4$ from the diagonal
curvature, so its absolute error follows that of the larger normal constants;
on graphene it deviates by $5.5\%$ from the symmetry-imposed value, against
$1.6\%$ for the simple-shear route, which we therefore take as the primary
estimate. All five systems satisfy the Born criteria of
Eq.~\eqref{eq:born}.

The central quantity is the ultralow principal-axis shear stiffness of Twin
T-graphene, approximately $16$-fold below octa-graphene and $47$-fold
below graphene, whereas its axial stiffness $C_{11}=178.2$~N\,m$^{-1}$ remains
comparable to conventional two-dimensional semiconductor monolayers
\cite{liu2016mechanical,androulidakis2018tailoring}. The material is not
uniformly compliant: it stays stiff under axial loading and becomes
exceptionally soft under principal-axis shear.

This separation of scales is a property of the stiffness tensor itself, and
the Kelvin representation states it in the language of extremal elasticity.
For Twin T-graphene the eigenvalues are
$(\lambda_{\mathrm{comp}},\lambda_{\mathrm{dev}},\lambda_{\mathrm{sh}})
=(176.8,\ 179.6,\ 6.06)$~N\,m$^{-1}$: two large normal eigenmodes and one
small shear eigenmode, the signature of a unimode-like response
\cite{milton1995elasticity}, with $\mathcal{R}_u=29.2$. The remaining systems
do not carry a comparably soft eigenvalue and range from nearly isotropic to
moderately anisotropic, with $\mathcal{R}_u=1.01$ for Me-graphene
$(203.9,\ 206.8,\ 210.7)$, $1.02$ for graphene $(285.6,\ 290.3,\ 423.9)$,
$1.14$ for penta-graphene $(248.2,\ 282.6,\ 304.2)$, and $2.52$ for
octa-graphene $(97.1,\ 244.3,\ 355.4)$~N \, m$^{-1}$. Twin T-graphene is thus
the only member of the set whose Kelvin spectrum isolates a single soft
eigenmode, and it does so with a small but finite $\lambda_{\mathrm{sh}}$: a
near-unimode covalent crystal rather than a mechanism carrying an exact
zero-energy mode \cite{milton2013complete,kadic2012practicability}.

The same hierarchy governs the in-plane anisotropy. The ratio
$(C_{11}-C_{12})/(2C_{66})$, which equals unity for an isotropic sheet, falls
between $0.93$ and $2.51$ for the other four systems, while Twin T-graphene
gives $29.6$, with $A_G=29.7$. It satisfies the first condition of
Eq.~\eqref{eq:isotropy} and violates the second by more than an order of
magnitude, which is why the equality of the axial stiffnesses does not
establish elastic isotropy and revises the previously proposed isotropic
description of this monolayer \cite{bhattacharya2020twin}.

Figure~\ref{fig:polar} translates the elastic constants into their directional representations, in which an isotropic response produces a circle. Graphene, penta-graphene, and Me-graphene trace nearly circular curves in all panels, while octa-graphene shows a moderate four-lobed modulation. Twin T-graphene is qualitatively different: $Y(\theta)$ falls from
$178.1$~N\,m$^{-1}$ along the lattice axes to $11.7$~N\,m$^{-1}$ at
$45^{\circ}$, a 15-fold variation, while $G(\theta)$ falls from
$89.8$~N\,m$^{-1}$ along the diagonal to $3.03$~N\,m$^{-1}$ along the principal axes. Its Poisson's ratio spans $-0.008$ along the axes to $0.93$ at
$45^{\circ}$, approaching the value $\nu=1$ at which an in-plane isotropic
sheet becomes areally incompressible, consistent with a mechanism dominated by
internal sublattice reorientation rather than by uniform C--C bond stretching.
The spread across the set is not organized by the point group either:
Me-graphene and penta-graphene share $D_{2d}$ and are both nearly isotropic, while octa-graphene and Twin T-graphene share $D_{4h}$ and differ by more
than an order of magnitude, $A_G=2.5$ versus $29.7$. Two crystals with
identical symmetry-allowed tensor forms therefore realize entirely different
shear responses, which locate the origin of the effect in the internal
structure rather than in the space group.

\begin{figure*}[t]
\centering
\includegraphics[width=0.9\textwidth]{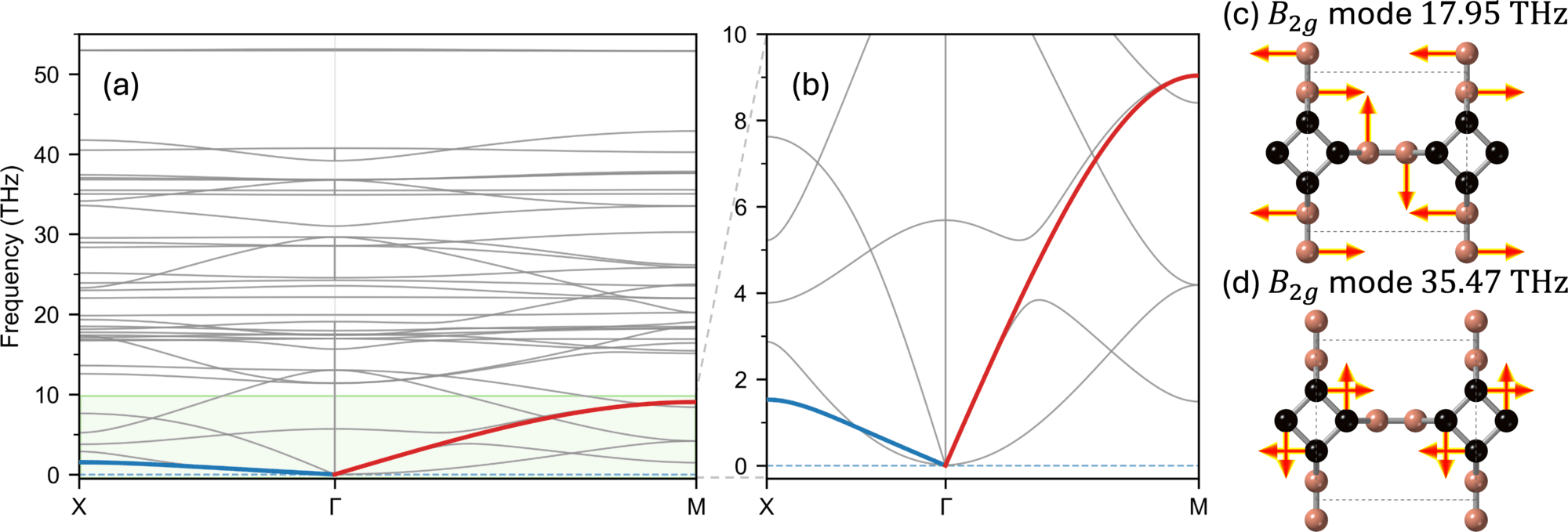}
\caption{Phonon validation of the ultralow shear stiffness.
\textbf{(a)} Phonon dispersion of Twin T-graphene along $X$--$\Gamma$--$M$; the shaded
area marks the window enlarged in (b).
\textbf{(b) }Low-frequency region. The in-plane transverse acoustic branch is
highlighted in blue along $\Gamma$--$X$ ($[100]$) and in red along
$\Gamma$--$M$ ($[110]$), assigned by polarization; the transverse sound
velocities are the slopes of these branches near $\Gamma$, and 
their ratio above five equals the square root of the static ratio $(C_{11}-C_{12})/(2C_{66})$.
\textbf{(c,d)} Zone-center $B_{2g}$ optical eigenvectors at $17.95$ and $35.47$~THz; arrows indicate the in-plane atomic displacements. These modes span the
internal subspace that couples linearly to shear, and their admixture into the
TA branch accompanies its flat dispersion along $\Gamma$--$X$.}
\label{fig:phonon_validation} 
\end{figure*}

\subsection{Origin of the ultralow shear stiffness}
\label{sec:shear_origin}

\subsubsection{Microscopic shear mechanism}

Relaxed geometries identify the coordinate that absorbs the deformation.
After removal of the imposed affine deformation, the maximum non-affine atomic
displacement at $|\gamma|=0.04$ is $0.026$~\AA, with a component normal to the
monolayer below $10^{-3}$~\AA, so the internal relaxation is almost entirely
in-plane and involves no appreciable additional buckling.

The dihedral angles associated with the mid-plane
$\mathrm{C_B}$--$\mathrm{C_B}$ bonds reveal how: for each mid-plane bond, the
dihedral involving the upper sublayer rotates by $+4.96^{\circ}$ while its
lower counterpart rotates by $-4.96^{\circ}$. The same antisymmetric pattern
occurs for each equivalent axis of the primitive cell, and reversing the
sign of $\gamma$ reverses all rotations. Macroscopic shear is thus
accommodated through a counter-torsion of the two tetragonal sublayers about
the mid-plane bonds: the $\mathrm{C_A}$ tetragonal units behave as nearly
rigid structural motifs, whereas the short $\mathrm{C_B}$--$\mathrm{C_B}$
bonds act as local torsional axes. The mechanism relies on the three-sublayer
architecture, which allows the covalent network to change its dihedral
configuration while preserving most local bond lengths and bond angles, and
which the strictly planar octa-graphene lacks.

\subsubsection{Hierarchy of constrained shear responses}
\label{sec:constrained}

The role of internal relaxation is quantified through two constraints. First,
the four mid-plane $\mathrm{C_B}$ atoms are held at their affinely deformed
positions while all remaining atoms relax; second, the complete atomic basis
is constrained to follow the affine deformation, giving the clamped-ion value.
The three calculations establish the hierarchy
\begin{equation}
C_{66}^{\mathrm{relaxed}} (3.03) < C_{66}^{\mathrm{C_B\text{-}fixed}} (18.06)
< C_{66}^{0} (81.43) \quad \text{in} ~\mathrm{N\,m^{-1}}.
\label{eq:shear_hierarchy}
\end{equation}
Therefore, full internal relaxation removes $96.3\%$ of clamped-ion shear
resistance, and the intermediate value shows that the response is cooperative:
the surrounding $\mathrm{C_A}$ framework contributes substantially on its own,
while releasing the mid-plane sublattice produces the final sixfold reduction.
The soft Kelvin eigenvalue is a relaxation effect, not a signature of weak
covalent bonding.

The symmetry of the associated internal driving force follows from the
 irreducible representations of the zone-center. The optical modes at $17.95$ and
$35.47$~THz both transform as $B_{2g}$ under $D_{4h}$, and a projection of the
strain-odd clamped-ion forces onto this two-mode subspace recovers more than
$99.999\%$ of their squared norm. The principal-axis shear therefore couples
almost entirely with the $B_{2g}$ sector associated with the counter-torsional
relaxation; a quantitative internal-coordinate model of this renormalization
is given in Sec. S-VI.

\subsubsection{Validation from transverse acoustic phonons}

The static prediction is tested against the long-wavelength transverse
acoustic (TA) phonons of Fig.~\ref{fig:phonon_validation}. Along the principal
and diagonal directions of the tetragonal lattice, the Christoffel equation
separates into purely longitudinal and transverse in-plane branches, with
\begin{equation}
\rho_{\mathrm{2D}}\,v_{\mathrm{TA}}^{2}([100]) = C_{66},
\qquad
\rho_{\mathrm{2D}}\,v_{\mathrm{TA}}^{2}([110]) = \frac{C_{11}-C_{12}}{2},
\label{eq:vta}
\end{equation}
where $\rho_{\mathrm{2D}}=7.964\times10^{-7}$~kg \, m$^{-2}$ is the density of the areal mass. The branches are assigned by polarization, which separates the
in-plane TA modes from the quadratic flexural branch.
The dispersion does not carry imaginary frequencies anywhere in the
Brillouin zone, and the low-frequency branches converge with
respect to the supercell size: from $3\times3\times1$ to $6\times6\times1$ changes the
transverse acoustic slope along $[100]$ by 1.0\% and every branch by less
than 0.04~THz, so the compliant channel is a soft but stable
mode rather than an incipient instability.
Equation~\eqref{eq:vta} predicts $v_{\mathrm{TA}}([100])=1.95$ and
$v_{\mathrm{TA}}([110])=10.62$~km\,s$^{-1}$, and the calculated slopes near
$\Gamma$ reproduce both within $10\%$ (Sec. S-VII). The two
branches differ in slope by more than a factor of five, and the $[100]$
velocity is the smallest transverse sound velocity in the plane of the structure:
an independent dynamical measurement of the same soft channel.

The acoustic branch and static relaxation, in addition, share the same internal
displacement pattern. Because an acoustic eigenvector at $\Gamma$ is a rigid
translation, the comparison is made at small finite wave vectors after
projecting out the translations, using the overlap of
Eq. S10: along $\Gamma$--$X$, $\mathcal{O}\simeq0.98$, with
the four mid-plane atoms contributing about $42\%$ of the modal displacement.
The two $B_{2g}$ frequencies are high, so $\lambda_{\mathrm{sh}}$ does not
originate from a softening optical mode but from the relaxed-ion cancellation quantified in Sec.~\ref{sec:criterion}; the flat blue dispersion of
Fig.~\ref{fig:phonon_validation}(b) is the dynamical imprint of the same
counter-torsional channel that softens the static shear constant.

\subsection{Electronic response to shear}
\label{sec:bands}

\begin{figure*}[!]
\centering
\includegraphics[width=0.99\textwidth]{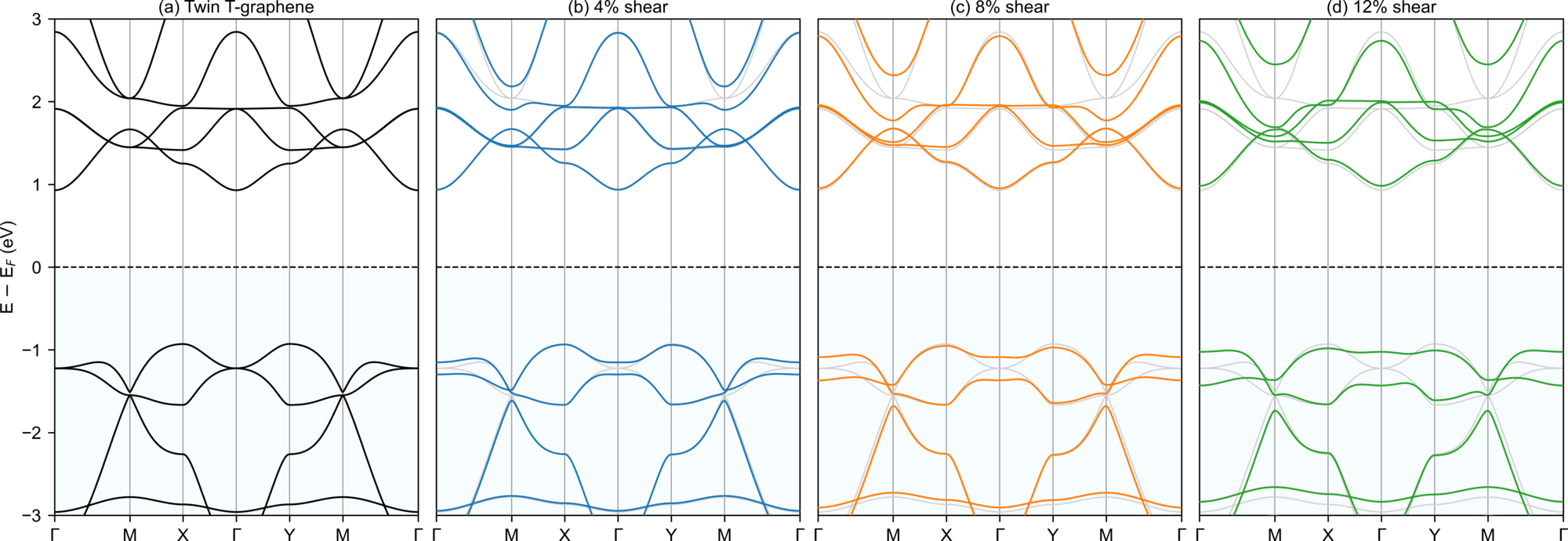}
\caption{
Electronic bands of Twin T-graphene under principal-axis simple
shear,
$\mathbf F_{\mathrm s}
=\mathbf I+\gamma\,\mathbf e_x\otimes\mathbf e_y$,
whose symmetric strain component transforms as $B_{2g}$.
Panel\textbf{ (a)} shows the undeformed structure, while panels \textbf{(b)--(d)}
correspond to $\gamma=0.04$, $0.08$, and $0.12$. Gray curves in
panels (b)--(d) reproduce the undeformed bands for reference.
All the energies are referred to the mid-gap level. For the sheared structures, $X$, $Y$, and $M$ denote the same fractional
reciprocal-space coordinates as in the undeformed cell; the $B_{2g}$ strain
removes the $C_4$ operation that relates $X$ and $Y$, so these points are no
longer symmetry equivalent.   
}
\label{fig:shear_bands}
\end{figure*}

Figure~\ref{fig:shear_bands} shows that compliance does not need to come at the
cost of the electronic structure. The $B_{2g}$ strain removes the $C_4$
rotation relating $k_x$ and $k_y$, and the valence manifold registers this
directly through a splitting of the states in $X$ and $Y$ and a separation of
the two uppermost occupied bands. The fundamental gap, in contrast, retains
its indirect character and increases only from $1.85$ to $1.95$~eV up to
$\gamma=0.12$, without semiconductor-to-metal transition. This follows from
the mechanism: shear is accommodated by counter-torsion about the mid-plane
bonds, so the covalent topology and the orbital overlaps that set the gap
survive while the near-edge valence states reorganize.

\section{Effective internal-coordinate model and design principles}
\label{sec:criterion}

The above results admit a compact relaxed-ion formulation. Writing the
harmonic shear energy in terms of the engineering shear strain $\gamma$ and the zone-center internal coordinates that transform as the same irreducible
representation as the applied strain, and minimizing over those coordinates,
gives the following,
\begin{equation}
C_{66} = C_{66}^{0} - \chi_{\mathrm{int}},
\label{eq:renormalization}
\end{equation}
where $C_{66}^{0}$ is the shear stiffness before relaxation of the active
internal subspace and $\chi_{\mathrm{int}}$ is an internal-strain
susceptibility built from the strain--mode couplings and the force-constant
matrix of that subspace (Sec. S-VI). An ultralow but
mechanically stable shear constant requires $\chi_{\mathrm{int}}$ to approach
$C_{66}^{0}$ from below, which turns the observed compliance into three
conditions that a candidate crystal must meet simultaneously.

\begin{enumerate}
\item[(i)] \emph{Symmetry-allowed strain--mode coupling.} An internal
displacement subspace must transform as the same irreducible representation as
the target strain, so that a bilinear coupling is allowed. The condition is
directly testable by classifying the odd non-affine displacement under
opposite strains and projecting it onto the zone-center phonon eigenvectors.

\item[(ii)] \emph{Large internal-strain susceptibility.} The correction 
$\chi_{\mathrm{int}}$ must approach $C_{66}^{0}$, through strong strain--mode
coupling, weak restoring forces, or both. Rotational and dihedral patterns are
favorable because they generate substantial internal motion while limiting
costly bond-length and bond-angle distortions.

\item[(iii)] \emph{A stiff complementary load-bearing network.} The motifs not
involved in the compliant coordinate must continue to resist the channels
intended to remain rigid, which is what separates mode-selective compliance
from uniform softness and preserves mechanical stability.
\end{enumerate}

Twin T-graphene satisfies the three conditions through the $B_{2g}$
counter-torsional subspace identified above, whose amplitude is linear in
strain and cannot be linearly coupled to the $A_{1g}$ and $B_{1g}$
normal-strain channels; this is the symmetry basis for the separation between
shear compliance and longitudinal stiffness. Evaluating the model parameters
from the calculated frequencies and the fitted slopes gives
$\chi_{\mathrm{int}}\simeq79$~N\,m$^{-1}$ and $C_{66}^{0}\simeq82$~N\,m$^{-1}$,
reproducing the clamped-ion value $81.43$~N\,m$^{-1}$ obtained independently
from the constrained relaxations of Sec.~\ref{sec:constrained}.

An out-of-family negative control shows that the conditions are not redundant.
The puckered, ferroelastic monolayer $\mathrm{PdS_2}$
\cite{zhang2021flexible}, whose corrugated pentagonal architecture with
chalcogen dimers might suggest analogous relaxation pathways, gives
$C_{66}/C_{11}\simeq0.30$ against $0.017$ for Twin T-graphene and a Kelvin
spectrum with no extremal separation (Sec. S-VIII); its ferroelasticity
involves switching between symmetry-equivalent minima across a finite barrier
rather than a continuous low-energy coordinate within a single phase, so
condition~(ii) fails despite pronounced corrugation. Corrugation, thickness,
and even proximity to ferroelastic switching are therefore insufficient
descriptors on their own. Equation~\eqref{eq:renormalization} also places the
crystal relative to architected pentamodes, in which finite joint stiffness
sets the residual shear resistance \cite{milton1995elasticity,
kadic2012practicability,zhou2024two}: here the analogous residual is fixed by
 competition between $C_{66}^{0}$ and $\chi_{\mathrm{int}}$, both set by
chemistry and atomic topology rather than by a geometrical joint parameter.


\section{Robustness beyond the harmonic limit}
\label{sec:robustness}

The elastic constants are harmonic quantities evaluated at $T=0$. Two
independent questions follow: whether the compliant channel survives a finite
strain and whether it survives a finite temperature.

\subsection{Nonlinear large-deformation response}
\label{sec:nonlinear}

Quasistatic deformation scans with MACE-MH-1 on a $4\times4$ supercell extend
the four loading paths well beyond the harmonic window
(Sec. S-IX-C). The potential reproduces the DFT
hierarchy $C_{11}>(C_{11}-C_{12})/2\gg C_{66}$ and the shear anisotropy, but
not the small residual $C_{66}$ to first-principles precision 
(Sec. S-IX), so these results are used for relative trends and
deformation pathways rather than absolute thresholds. Within that scope, the
four paths behave very differently. The principal-axis tension carries the highest
stress, $23.72$~N\,m$^{-1}$, and fails catastrophically at
$\varepsilon=0.240$, while the equibiaxial and diagonal loads pass through a
reconstructive precursor in which the mean coordination rises from $3.00$ to
$3.67$ and the layer flattens, fracturing later at $0.350$ and $0.470$. Simple
shear reaches only $6.07$~N\,m$^{-1}$ but sustains by far the largest
deformation: its first instability at $\gamma\simeq0.46$ involves no bond
formation or breaking and an energy change of only $-11$~meV per atom, so it
is a buckling instability and no catastrophic fracture occurs within the
investigated interval. The compliant shear channel therefore persists in the
nonlinear regime and accommodates large deformation through buckling and
reconstruction rather than immediate brittle failure.

\subsection{Finite-temperature stability}
\label{sec:aimd}

We perform Born--Oppenheimer molecular dynamics simulations on a $2\times2$
supercell (48 atoms) in the canonical ensemble at $300$~K, with a
Nos\'e--Hoover thermostat, a $1$~fs time step, and $20$~ps of simulation, for
the relaxed monolayer and for the same structure constrained at
$\gamma=0.12$, well beyond the interval used to extract $C_{66}$. Both remain
dynamically stable: every atom retains threefold coordination, the initial
bond network is fully preserved, and the aligned RMSD stays bounded, so
thermally activated bond breaking and progressive reconstruction are excluded
on this time scale (diagnostics in Sec. S-X). The MD supercell is too small to
represent long-wavelength flexural fluctuations, so the trajectories test the
integrity of the local bonding rather than the rippling behavior of an
extended membrane.

\begin{figure}[t]
\centering
\includegraphics[width=\columnwidth]{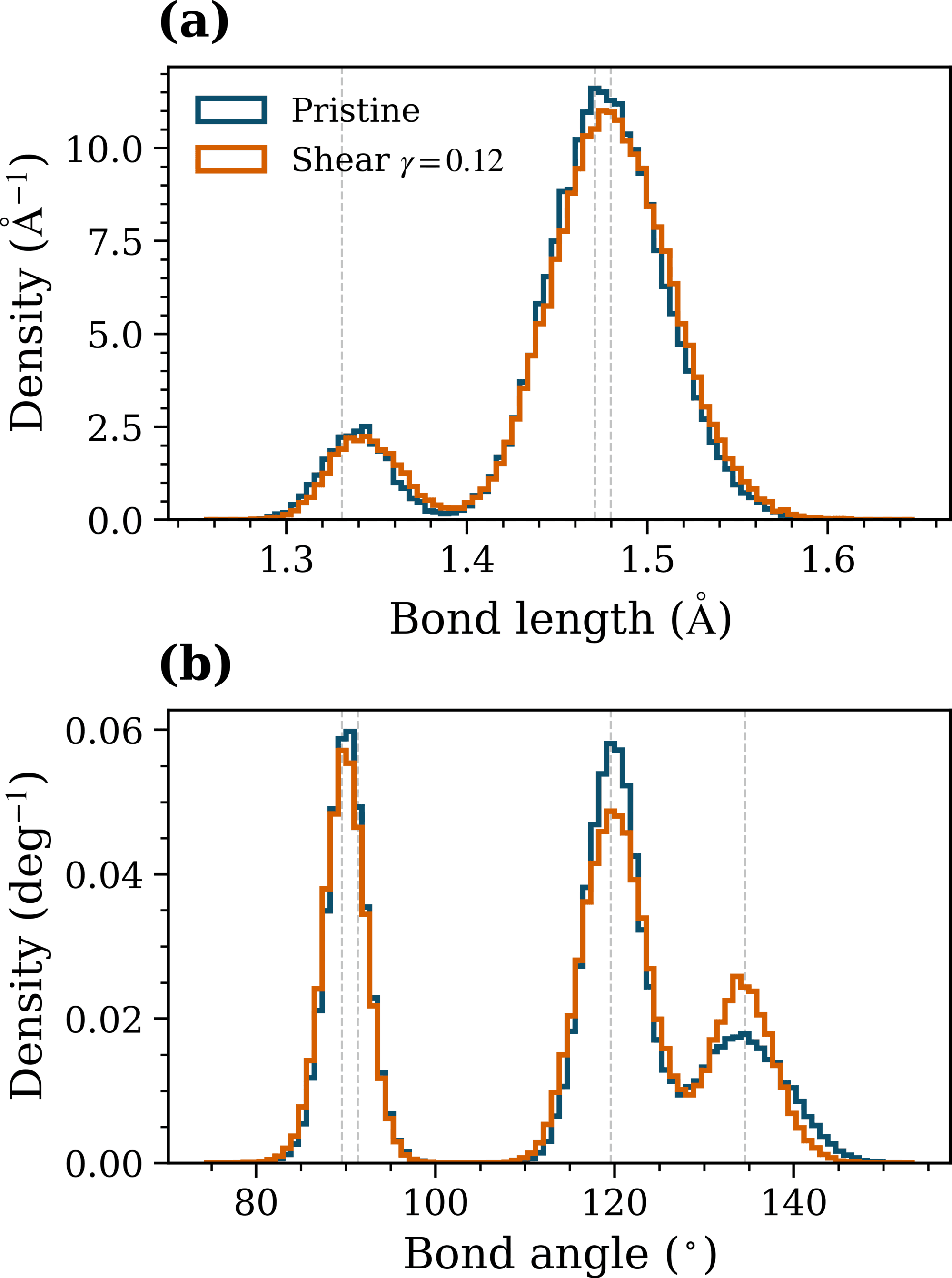}
\caption{
Finite-temperature local structure of pristine and sheared Twin T-graphene
from AIMD at $300$~K.
\textbf{(a)} Probability density of first-neighbor bond lengths for the
pristine monolayer and for the monolayer constrained at $\gamma=0.12$. The
nearly overlapping distributions show that the imposed shear produces
negligible bond stretching.
\textbf{(b)} Corresponding bond-angle probability densities. In contrast to
the bond lengths, shear redistributes the populations of the higher-angle
environments while leaving the tetragonal-ring populations near $90^\circ$
essentially unchanged.
Vertical dashed lines mark the bond lengths and bond angles of the relaxed
pristine structure at $t=0$, which are near the maxima of the
finite-temperature distributions.
}
\label{fig:aimd}
\end{figure}

Figure~\ref{fig:aimd} shows how the imposed deformation is accommodated
locally. The bond length distributions of the two trajectories are nearly
superimposed, the mean bond length changing by $0.1\%$, while the angular
distribution transfers weight from the environment near $120^\circ$ to the one
near $134^\circ$, and the tetragonal-ring populations near $90^\circ$ remain
unchanged. At $300$~K, as at $T=0$, shear is accommodated through reorientation
of the bridging geometry rather than through distortion of the rings. The
trajectories also give an energetic check: the sheared run exceeds the
pristine one by $\Delta\langle E_{\mathrm{pot}}\rangle=3.81\pm0.35$~meV per
atom, compared to the static strain energy $\tfrac{1}{2}C_{66}\gamma^{2}A_0/N =
3.41$~meV per atom at $T=0$, since the harmonic thermal
contribution cancels in the difference. Recasting through the same quadratic relation, this gives an effective energetic stiffness
$C_{66}^{E}(300~\mathrm{K}) = 3.38\pm0.31$~N\,m$^{-1}$, close to the $T=0$
value \footnote{$C_{66}^{E}$ is an energetic estimate at fixed global strain,
not a thermodynamic isothermal elastic constant, which would require a
free-energy derivative or an explicit finite-temperature stress--strain
calculation.}. Therefore, a $12\%$ shear remains thermally accessible without
compromising the topology or the local bonding of the monolayer.

\section{Outlook: candidate materials}
\label{sec:outlook}

The design variable identified here is not low stiffness or porosity, but the
coexistence of a rigid load-bearing framework with a symmetry-compatible
internal coordinate that couples to the target strain while avoiding
substantial bond stretching. The most interesting consequence is that a higher
symmetry should change the outcome qualitatively rather than reproduce it.
Twin graphene \cite{jiang2017twin} shares the three-sublayer architecture that
supplies the compliant coordinate here, but in its $P6/mmm$ structure the two
in-plane deviatoric strains form the $E_{2g}$ doublet rather than two separate
representations, so coupling to an internal $E_{2g}$ sector would renormalize
both shear eigenvalues equally while preserving long-wavelength isotropy. If
both fell well below the compressional eigenvalue, the crystal would approach
the two-dimensional bimode analog of a pentamode solid, a qualitatively
different regime from the unimode-like response found here; testing this
requires an explicit clamped-ion and relaxed decomposition of its shear
response, as do the wider tetragonal allotrope family
\cite{ghosal2022beyond,majidi2024twin48,fabris2020new} and two-dimensional
covalent organic frameworks \cite{xiong2025cofmechanics,thakur2023thermalcof,
auras2024dynamiccof}, where linker torsion offers a chemically programmable
route to the same coordinate.

The most direct experimental signature is acoustic. The contrast between
$v_{\mathrm{TA}}([100])=1.95$ and $v_{\mathrm{TA}}([110])=10.62$~km\,s$^{-1}$
implies direction-selective propagation and polarization of in-plane elastic
waves without any patterned architecture, and it is measurable by inelastic
scattering or Brillouin spectroscopy on a supported monolayer.

\section{Final Remarks}
\label{sec:conclusions}

Four tetragonal carbon monolayers possessing the same symmetry-allowed elastic tensor form display a more than 50-fold variation in shear stiffness, whereas their axial stiffnesses differ by less than a factor of two. Twin T-graphene exhibits the most pronounced contrast, with $C_{66}=3.03$~N\,m$^{-1}$ and a Kelvin spectrum in which one soft shear eigenvalue is a factor of 29.2 lower than the two normal-strain eigenvalues, while the Born stability criteria are satisfied. Consequently, a densely bonded covalent crystal can approach the unimode limit without compliant hinges, porosity, or engineered mesoscale architecture.

The soft channel arises from internal relaxation rather than weak bonding. Symmetry confines the linear shear coupling to the $B_{2g}$ sector, and the calculated non-affine displacements lie within this sector to numerical accuracy. The resulting counter-torsion of the tetragonal sublayers about the mid-plane bonds eliminates 96.3\% of the clamped-ion shear stiffness, while bond-length changes remain below 0.15\% at $|\gamma|=0.04$. An internal-coordinate model reproduces this cancellation quantitatively. The soft channel predicts transverse sound velocities of 1.95~km\,s$^{-1}$ along [100] and 10.62~km\,s$^{-1}$ along [110], providing a testable acoustic signature. The structure remains dynamically stable and maintains its local bonding at 300~K under finite shear, and the compliant pathway persists beyond the harmonic regime. Since the covalent topology is preserved during deformation, the semiconducting gap also remains open, with its indirect value increasing only from 1.85 to 1.95~eV up to $\gamma=0.12$.

These findings show that symmetry determines the allowed relaxation channels, while bonding topology and strain--mode coupling govern the extent of elastic softening. This mechanism indicates three criteria for identifying similar behavior in other crystals: the presence of a symmetry-compatible internal displacement subspace, an internal-strain susceptibility approaching the corresponding clamped-ion stiffness, and a complementary bonding network that remains rigid under other deformation modes. Symmetry-resolved relaxations and comparisons between relaxed-ion and clamped-ion elastic responses are sufficient to evaluate these criteria. Although demonstrated here for a single material, this framework offers a microscopic basis for identifying extreme mode-selective elasticity in covalent crystals without engineered mechanical mechanisms.

\paragraph*{Supplemental Material.}
The Supplemental Material provides additional details on
(i) energy--strain fitting, elastic-constant extraction,
and comparisons with published elastic data;
(ii) directional elastic moduli, bond-length and bond-angle
responses, and symmetry-resolved mode projections;
(iii) the internal-coordinate model and transverse acoustic
velocities; and (iv) the $\mathrm{PdS_2}$ benchmark,
small-strain validation of the machine-learning interatomic
potential and its application to large-deformation simulations,
and finite-temperature {ab initio} molecular-dynamics
analysis. The Supplemental Material also cites Refs.~\cite{lee2008measurement,
cong2019probing,wang2023mech2d,qin2015phosphorene,
gao2025elastic,mannan2025uniffbench}.

\paragraph*{Acknowledgments:}
JWG and MP acknowledge financial support from ANID-FONDECYT (Chile) grant No. 1262499.
Powered@NLHPC: This research was partially supported by the NLHPC supercomputing infrastructure (CCSS210001).

\bibliography{suya}

\end{document}